\documentclass{ws-ijmpd}

\begin{document}

\markboth{Marco Frasca}
{Strong coupling expansion for general relativity}

\catchline{}{}{}{}{}
%
%%%%%%%%%%%%%%%%%%%%%%%%%%%%%%%%%%%%%%%%%%%%%%%%%%%%%%%%%%%%%%%%%%%%

\title{Strong coupling expansion for general relativity}

\author{MARCO FRASCA}

\address{Via Erasmo gattamelata, 3\\
00176 Roma (Italy)\\
marcofrasca@mclink.it}

\maketitle

%\begin{history}
%\received{Day Month Year}
%\revised{Day Month Year}
%\comby{Managing Editor}
%\end{history}

\begin{abstract}
Strong coupling expansion is computed for the Einstein equations in vacuum in the 
Arnowitt-Deser-Misner (ADM) formalism. The
series is given by the duality principle in perturbation theory as presented in 
[M.Frasca, Phys. Rev. A {\bf 58}, 3439 (1998)]. An example of
application is also given for a two-dimensional model of gravity expressed through the Liouville
equation showing that the expansion is not trivial and consistent with the exact solution,
in agreement with the general analysis. Application to the Einstein equations in vacuum in 
the ADM formalism shows that the spacetime near singularities is driven by 
space homogeneous equations. 
\end{abstract}


\section{Introduction}	

Strong coupling expansions in theoretical physics have a longstanding tradition. One should
consider fluid dynamics where two different regimes of motions were found depending
on what term of the Navier-Stokes equation is taken as a perturbation \cite{flu}. Boundary
layer problems in perturbation theory firstly appeared in this way \cite{Nay}. In the 
seventies attempts to build a strong coupling expansion for quantum field theory were put
forward by Bender and colleagues \cite{bend1,bend2,bend3}. These latter expansions faced the problem
of the dependence on a cut-off that was not possible to dispose of in a well definite
manner. The idea behind was simple
and smart: For a self-interacting scalar field the perturbation is not the potential but the kinetic term.

What makes these methods rather interesting is that they exploit an evident symmetry in the
choice of the perturbation in a differential equation. This choice is arbitrary and different
perturbation schemes can be devised depending on it. So, one may ask what is
the relationship between different series. This matter was exploited in Ref.\refcite{fra1}
where a theorem was proved showing that in the strong coupling expansion for a quantum system
the leading order is obtained by the adiabatic theorem in quantum mechanics. This striking
result was further analyzed in Ref.\refcite{fra2} showing how the proper expansions can be obtained
in different regimes for some Hamiltonians in quantum optics. A duality principle holds that
interchanging the terms in the perturbation series means to obtain a series with the
reversed ordering parameter. A successful application of this idea in quantum field theory
has been recently given \cite{fra3}.

The leading order is given by the adiabatic theorem in quantum mechanics for the Schr\"odinger
equation but this result is indeed ubiquitous and we will show it as a generic result for
a strong coupling expansion for a differential equation. This theorem will permit us to
extend the above analysis to Einstein equations in vacuum in the form given in the 
Arnowitt-Deser-Misner (ADM) formalism \cite{adm}. Indeed,
a strongly perturbed system, understood as a system undergoing a largely increasing perturbation,
becomes a system with an increasingly slow dynamics, that is, an adiabatic system. More
importantly for our aims is the fact that these equations behave homogeneously with
respect to the spatial variable in a regime that appears dual, in the sense of
perturbation theory \cite{fra1}, to the case of the gravitational waves that represent
small perturbations on a given spacetime.

The relevance of this result for the Einstein equations is easily understood being open up 
several problems that presently only a numerical treatment can manage with all the shortcomings that
this implies \cite{alc,berg}. One of the questions that could benefit from a strong coupling 
analysis is the singularity problem \cite{frie}. Indeed, it is well-known that while
theorems exist due to Penrose and Hawking \cite{pen,haw,penhaw} proving the existence of
singularities in the solutions of the Einstein equations, but very few is known about the nature
of such solutions and their behavior near the singularity itself. Our approach permits
to analyze the Einstein equations in a regime of very intense gravitational field and then
to get an answer to this problem. Indeed, our approach permits to conclude that the dynamics near 
the singularity is described by homogeneous
solutions of the Einstein equations and some numerical
results hint in this direction \cite{berg,garf}. 

What we are going to show is that a dual expansion exists to the weak field one that is commonly
used to study gravitational radiation\cite{wald}. The main result we obtain in this way is
that the leading order equations are just the homogeneous part of the Einstein equations.
This reduces the spatial dependence to parameters in the integration constants and
the gauge choice at this order. This in turn means that in a regime of strong gravitational field
the solutions to be considered are the homogeneous ones. We
will see that in two dimensions the situation is quite similar producing a non-singular
solution but with an analogous duality between
weak and strong field solutions. As a by-product we will obtain
a strong coupling expansion for the gravity in two dimensions as
described by the Liouville equation\cite{teit,jack1,jack2}. We will show the agreement with the
exact solution giving full consistency to our approach for a well known non-linear case. 
This will yield a start up to our perturbation procedure toward the final proof.

The paper is so structured. In sec.\ref{sec2} we present duality in perturbation theory and
the adiabatic approximation at the leading order. In sec.\ref{sec3} the approach, in a
general ADM-like setting, is applied to two dimensional gravity. In sec.\ref{sec4} the
method is applied to Einstein equations in vacuum in the ADM formalism. 
In sec.\ref{sec5a} we give some applications showing the $1/G$ behavior of the
perturbation series. Finally, in sec.\ref{sec5} conclusions are given.

\section{Duality in perturbation theory and adiabatic approximation\label{sec2}}

In this section we will show how a duality principle holds in perturbation theory
showing how to derive a strong coupling expansion with the leading order ruled by
an adiabatic dynamics in order to study the evolution of a physical system. 
We consider the following perturbation problem
\begin{equation}
\label{eq:eq1}
    \partial_t u = L(u) + \lambda V(u)
\end{equation}
being $\lambda$ an arbitrary ordering parameter: As is well known
an expansion parameter is obtained by the computation of the series itself. The standard 
approach assume the limit $\lambda\rightarrow 0$ and putting
\begin{equation}
    u = u_0 + \lambda u_1 +\ldots
\end{equation}
one gets the equations for the series
\begin{eqnarray}
    \partial_t u_0 &=& L(u_0) \\ \nonumber 
    \partial_t u_1 &=& L'(u_0)u_1 + V(u_0) \\ \nonumber 
    &\vdots&
\end{eqnarray}
where a derivative with respect to the ordering parameter is indicated by a prime. We recognize here a conventional
small perturbation theory as it should be. But the ordering parameter is just a conventional matter and so one
may ask what does it mean to consider $L(u)$ as a perturbation instead with respect to the same parameter.
Indeed one formally could write the set of equations
\begin{eqnarray}
\label{eq:set}
    \partial_t v_0 &=& V(v_0) \\ \nonumber 
    \partial_t v_1 &=& V'(v_0)v_1 + L(v_0) \\ \nonumber 
    &\vdots&
\end{eqnarray}
where we have interchanged $L(u)$ and $V(u)$ and renamed the solution as $v$. The question to be answered is
what is the expansion parameter now and what derivative the prime means. To answer this question we rescale the
time variable as $\tau = \lambda t$ into eq.(\ref{eq:eq1}) obtaining the equation
\begin{equation}
\label{eq:eq2}
    \lambda\partial_{\tau} u = L(u) + \lambda V(u)
\end{equation}
and let us introduce the small parameter $\epsilon=\frac{1}{\lambda}$. It easy to see that applying again the
small perturbation theory to the parameter $\epsilon\rightarrow 0$ we get the set of equations (\ref{eq:set}) 
but now the time is scaled as $t/\epsilon$, that is, at the leading order the development parameter of the
series will enter into the scale of the time evolution producing a proper slowing down ruled by the equation
\begin{equation}
\label{eq:lead}
    \epsilon\partial_t v_0 = V(v_0)
\end{equation}
that we can recognize as an equation for adiabatic evolution that in the proper limit $\epsilon\rightarrow 0$
will give the static solution $V(u_0)=0$. We never assume this latter solution but rather we will study
the evolution of eq.(\ref{eq:lead}). Finally, the proof is complete as we have obtained a dual series
\begin{equation}
    u = v_0 + \frac{1}{\lambda} v_1 +\ldots
\end{equation}
by simply interchanging the terms for doing perturbation theory. This is a strong coupling expansion
holding in the limit $\lambda\rightarrow\infty$ dual to the small perturbation theory $\lambda\rightarrow 0$
we started with and having an adiabatic equation at the leading order. 

It is interesting to note that, for a partial differential equation, 
we can be forced into a homogeneous equation because, generally, if we require
also a scaling into space variables we gain no knowledge at all on the evolution of a
physical system. On the other side, requiring a scaling on the space variables and not on
the time variable will wash away any evolution of the system. So, on most physical systems
a strong perturbation means also a homogeneous solution but this is not a general rule. As
an example one should consider fluid dynamics where two regimes dual each other can be found
depending if it is the Eulerian or the Navier-Stokes term to prevail. In general relativity
things stay in a way to get a homogeneous equation at the leading order. The reason for this
is that products of derivatives or second order derivatives in space coordinates are
the only elements forming the Einstein tensor beside time dependence.

This completes the analysis initially realized in quantum mechanics in Ref.\refcite{fra1,fra2} 
permitting a direct application to general relativity.

\section{Strong coupling expansion for two-dimensional gravity\label{sec3}}

As is well-known a direct computation of the Ricci tensor in two dimensions gives zero proving
that two-dimensional general relativity is trivial. This is due to conformal invariance of the
theory in two dimensions. Different proposals have appeared to have non trivial solutions and
one of the most common theories is offered by the Liouville model \cite{teit,jack1,jack2}
that has had a lot of applications also in string theory\cite{tes}. Our aim in this
section is to obtain a dual perturbation series
% Modified on 10-10-2005
to the weak perturbation series as given in \cite{jack1,jack2}
% end mod
for this model of two dimensional gravity
in order
% Modified on 10-10-2005
just
% end mod  
to give a scheme of working of the machinery to be also applied to the full
four dimensional case. So, let us write the general line element in two dimensions in the
form proper to ADM formalism
\begin{equation}
    ds^2 = [-\alpha(x,t)^2+\beta_1(x,t)^2]dt^2+2\beta_1(x,t)dtdx+\gamma(x,t) dx^2
\end{equation}
where we have set $\alpha(x,t)$ for the lapse function, $\beta_1(x,t)$ for the shift function
and $\gamma(x,t)$ the metric factor in one dimension. We can simplify the metric by taking
the shift factor to be zero, so in the following $\beta_1(x,t)=0$ and this gives
\begin{equation}
    ds^2 = -\alpha(x,t)^2dt^2+\gamma(x,t) dx^2
\end{equation}
that for the equation $R=\Lambda$, with $R$ the Ricci scalar and $\Lambda$ a constant, yields
the ADM equations
\begin{eqnarray}
\label{eq:2dset}
    \partial_t \gamma &=& -2\alpha K \\ \nonumber
    \partial_t K &=& -\frac{\alpha}{\gamma}K^2 - \partial^2_x\alpha
    +\frac{1}{2\gamma}\partial_x\alpha\partial_x\gamma
    +\frac{\Lambda}{4}\alpha\gamma
\end{eqnarray}
with the freedom to fix the gauge function $\alpha$ to compute $K$,
the extrinsic curvature, and $\gamma$. A natural choice 
in this case is to put $\alpha^2 =\gamma$ to make explicit the conformal invariance of the metric
but other choices are also possible.
This gives back the Liouville equation \cite{teit,jack1,jack2}
\begin{equation}
    \Box\phi + \Lambda e^\phi = 0
\end{equation}
with the introduction of the field $\phi=\ln\gamma$ and being $\Lambda$ the only parameter 
of the theory. Indeed, a small perturbation theory is obtained by the explicit solution\cite{jack1}
\begin{equation}
    \phi(x,t) = \phi_0(x,t)-2\ln\left(1+\frac{\Lambda}{2}\int dx_1dt_1G(x-x_1,t-t_1)e^{\phi(x_1,t_1)}\right)
\end{equation}
being $\phi_0$ a solution of the equation $\Box\phi_0 = 0$ and $G$ can be expressed by the
functions $g(z)=\theta(z)$ or $g(z)=\theta(-z)$ with light cone coordinates as
$G(x-y)=\frac{1}{2}g(x^+-y^+)g(x^--y^-)$. As one can see in this case the start-up solution is a
solution of the free theory.

Eqs.(\ref{eq:2dset}) can be immediately used with the perturbation method of sec.\ref{sec2}
but this application is not so straightforward as one could think. There is a subtle point to
take into account as also will happen for the full dimensional case. Indeed, we note that
the lapse function $\alpha$ and the metric factor $\gamma$ are both without physical
dimensions so the extrinsic curvature $K$ has dimension $\left[\frac{1}{T}\right]$
as the derivative. This means that when we rescale time variable, we have also to
rescale $K$ properly. As the variable of our expansion is $\Lambda$ we see that we have to
do the following scale changing
\begin{eqnarray}
    t&\rightarrow&\sqrt{\Lambda}t \\ \nonumber
	  K&\rightarrow&\sqrt{\Lambda}K 
\end{eqnarray}  
but we can take into account of the rescaling in the extrinsic curvature in its series
development as
\begin{equation}
    K = \sqrt{\Lambda}\left(K_0 + \frac{1}{\Lambda}K_1 + \ldots\right).
\end{equation}
As said we also assume the gauge to be fixed with the conformal choice $\alpha^2=\gamma$.
With the substitution $\pi = -K/\sqrt{\gamma}$ the ADM equations become
\begin{eqnarray}
    \partial_t\alpha &=& \alpha\pi \\ \nonumber
	  \partial_t\pi &=& -\frac{\Lambda}{2}\alpha^2 + 
	  \partial_x\left(\frac{\partial_x\alpha}{\alpha}\right)
\end{eqnarray}
from which is straightforwardly obtained the Liouville equation taking into account that
$\alpha=e^{\frac{\phi}{2}}$. The Liouville equation is exactly solvable by two
arbitrary functions that depend on light cone coordinates \cite{teit,jack1,jack2}.  
It also follows the following set of perturbation equations
\begin{eqnarray}
    \partial_{\tau}\alpha_0 &=& \alpha_0\pi_0 \\ \nonumber
	  \partial_{\tau}\alpha_1 &=& \alpha_0\pi_1 + \alpha_1\pi_0 \\ \nonumber
	  \partial_{\tau}\alpha_2 &=& \alpha_0\pi_2 + \alpha_1\pi_1 + \alpha_2\pi_0 \\ \nonumber
                            &\vdots&  \\ \nonumber
    \partial_{\tau}\pi_0 &=& -{1 \over 2}\alpha_0^2 \\ \nonumber
	  \partial_{\tau}\pi_1 &=& -\alpha_0\alpha_1 + 
	  \partial_x\left(\frac{\partial_x\alpha_0}{\alpha_0}\right) \\ \nonumber
	  \partial_{\tau}\pi_2 &=& -\alpha_0\alpha_2 -\frac{1}{2}\alpha_1^2 +
    \partial^2_x\left(\frac{\alpha_1}{\alpha_0}\right) \\ \nonumber
	                     &\vdots&
\end{eqnarray}
being $\tau=\sqrt{\Lambda}t$ and we have adopted for $\pi$ the same scaling of the extrinsic curvature $K$
and we have considered $\alpha = \alpha_0 + \frac{1}{\Lambda}\alpha_1 + \ldots$.
At the leading order these equations give the homogeneous Liouville equation as it should be expected
\begin{equation}
    \partial^2_{\tau}\phi_0+e^{\phi_0}=0
\end{equation} 
that has the solution
\begin{equation}
\label{eq:2Dsol}
    \phi_0(\tau) =\ln\left[
	\frac{1}{2C_2^2\cosh^2\left(\frac{\tau+C_1}{2C_2}\right)}
	\right]  
\end{equation} 
so that $\alpha_0=e^{\frac{\phi_0}{2}}$. $C_1$ and $C_2$ are two integration constants that 
can depend on $x$ but not on $\Lambda$
otherwise the perturbation series would lose meaning. When $C_1$ and $C_2$ are taken not
depending on $x$ eq.(\ref{eq:2Dsol}) is an exact solution to the Liouville equation as can also be
proved taking the general solution to this equation \cite{jack1}
and expressing it by light cone coordinates. Then, higher order corrections are not zero
when e.g. $C_1$ is taken to be dependent on $x$. Assuming for the sake of simplicity
$C_1=kx$, being $k$ a constant, the equation to solve is
\begin{equation}
    \frac{d^2y}{d\tau^2}+\frac{y+k^2/2}{2C_2^2\cosh^2(\frac{\tau+kx}{2C_2})}=0
\end{equation}
being $y=e^{\frac{\phi_1-\phi_0}{2}}$ whose solution is trivially given by  $y=-k^2$ that is 
\begin{equation}
     \alpha_1=e^{\frac{\phi_1}{2}} = -\frac{k^2}{2}e^{\frac{\phi_0}{2}}
\end{equation}
and to second order 
\begin{equation}
     \alpha_2=e^{\frac{\phi_2}{2}} = -\frac{k^4}{8}e^{\frac{\phi_0}{2}}.
\end{equation}
We see that one has a series in $k^2/\Lambda$ as it should be in agreement with the exact
solution given by
\begin{equation}
     \alpha=\sqrt{1-\frac{k^2}{\Lambda}}e^{\frac{\phi_0}{2}}.
\end{equation}
and at the end of computation we can fix the constant $C_2$ as to have the scale factor 
$\sqrt{1-\frac{k^2}{\Lambda}}$ as a renormalization factor giving again an exact solution
\begin{equation}
\label{eq:2Dsolex}
    \phi(x,\tau) =-2\ln\left[
	\cosh\left(\frac{\sqrt{\Lambda} t+kx}{\sqrt{2}\sqrt{1-\frac{k^2}{\Lambda}}}\right)
	\right].  
\end{equation} 
We see that the ground state limit $\phi\rightarrow -\infty$ corresponds to the
dispersion relation $k^2=\Lambda$ producing a null metric. Besides, the dual weak 
coupling limit $k^2\gg\Lambda$ gives a ``gravitational wave''-like solution in a
complete analogy with the four dimensional case as we will see.

Finally, we have effectively got a series expansion in the parameter $\frac{1}{\Lambda}$ dual
to the small perturbation theory and in agreement with our adiabatic hypothesis in sec.\ref{sec2}.
The leading order is ruled by the homogeneuos Liouville equation. Besides, our approach led
us directly to exact solutions of the equation itself.

Some interesting conclusions can be drawn from the ``cosmological'' solution (\ref{eq:2Dsol}).
The metric factor is given by
\begin{equation}
\label{eq:2Dmf}
    \gamma(t) =\frac{1}{2C_2^2\cosh^2\left(\frac{\sqrt{\Lambda} t+C_1}{2C_2}\right)}  
\end{equation} 
that has the property to become a constant in the limit $\tau\rightarrow 0$ and this situation
is quite different from the full dimensional theory where from Kasner solution one can see
that metric blows up. This conformal factor becomes zero in the opposite limit $\tau\rightarrow\infty$
but generally our strong coupling expansion should have a proper time scale given by the
rescaling parameter $\Lambda$. As this parameter becomes larger also the time scale we
are able to analyze increases. In the present case we are in the favorable situation that the
perturbation terms contain global informations. Finally let us point out that such a kind of
cosmological solutions was found in string theory \cite{lmpx}.

\section{Strong coupling expansion for Einstein equations\label{sec4}}

Our approach for the case of Einstein equations in vacuum will be on the same lines as for the
two dimensional case. We would like to point out that the ground to reach our aim is derived 
from numerical general relativity as devised e.g. in Ref.\refcite{ct}. This means that we
consider Einstein equations in the ADM formalism in
a coordinate basis explicitly given by the set \cite{ct}
\begin{eqnarray}
    \partial_t\gamma_{ij}-\beta^l\partial_l\gamma_{ij} &=& \gamma_{lj}\partial_i\beta^l+
    \gamma_{il}\partial_j\beta^l-2\alpha K_{ij} \\ \nonumber
    \partial_tK_{ij}-\beta^l\partial_lK_{ij} &=& K_{il}\partial_j\beta^l+K_{jl}\partial_i\beta^l
    -2\alpha K_{il}K_j^l+\alpha K K_{ij} \\ \nonumber
    & &-{1 \over 2}\alpha\gamma^{lm}\left\{\partial_l\partial_m\gamma_{ij}+\partial_i\partial_j\gamma_{lm}
    -\partial_i\partial_l\gamma_{mj}-\partial_j\partial_l\gamma_{mi}\right. \\ \nonumber
    & & +\gamma^{np}\left[(\partial_i\gamma_{jn}+\partial_j\gamma_{in}
    -\partial_n\gamma_{ij})\partial_l\gamma_{mp}\right. \\ \nonumber
    & & +\partial_l\gamma_{in}\partial_p\gamma_{jm}-\partial_l\gamma_{in}\partial_m\gamma_{jp}\left.\right] \\ \nonumber
    & &-{1 \over 2}\gamma^{np}\left[(\partial_i\gamma_{jn}+\partial_j\gamma_{in}
    -\partial_n\gamma_{ij})\partial_p\gamma_{lm}+
    \partial_i\gamma_{ln}\partial_j\gamma_{mp}\left.\right]\right\} \\ \nonumber
    & &-\partial_i\partial_j\alpha+{1\over 2}\gamma^{lm}(\partial_i\gamma_{jm}+\partial_j\gamma_{im}
    -\partial_m\gamma_{ij})\partial_l\alpha
\end{eqnarray}
for a metric
\begin{equation}
    ds^2 = (-\alpha^2 + \gamma_{ij}\beta^i\beta^j)dt^2+2\beta^idx_idt+\gamma_{ij}dx^idx^j
\end{equation}
being $\alpha$ the lapse function, $\beta^i$ the shift vector and $\gamma_{ij}$ the spatial part of the metric.
The gauge freedom corresponds to fixing $\alpha$ and $\beta^i$ while $\gamma_{ij}$ is to be considered a fundamental
variable of the theory. $K_{ij}$ is the extrinsic curvature also a fundamental variable of the theory to be solved
for. Four constraint equations also hold
\begin{eqnarray}
     \ ^{(3)}R + K^2 - K_{ij}K^{ij} = 0 \\ \nonumber
     \ ^{(3)}\nabla_j (K^{ij}-\gamma^{ij}K) = 0
\end{eqnarray}
being $\ ^{(3)}R$ the Ricci scalar for $\gamma^{ij}$ and $\ ^{(3)}\nabla_j$ the corresponding covariant derivative.
We do not worry about the constraint equations in the following while these are relevant matter in numerical
computations. 

On the basis of the
duality principle in perturbation theory discussed above, we want to prove that in
a strong gravitational field the spatial part of the Einstein equations, at the leading order,
can be neglected and this behavior is exactly dual to the case of the gravitational
radiation in weak field expansion, a situation modeled in a very simple form in the case of
the Liouville equation. 

Our analysis starts from a small perturbation theory that can be derived 
from the ADM equations by fixing the gauge freedom as e.g.
\begin{eqnarray}
    \alpha &=& 1 \\ \nonumber
    \beta^i &=& 0.
\end{eqnarray}
Then we take
\begin{eqnarray}
    \gamma_{ij} &=& \eta_{ij} + \lambda \gamma_{ij}^{(1)} + + \lambda^2 \gamma_{ij}^{(2)} + \ldots \\ \nonumber
    K_{ij} &=& K_{ij}^{(0)} + \lambda K_{ij}^{(1)} + \lambda^2 K_{ij}^{(2)} + \ldots
\end{eqnarray}
being $\eta_{ij}=\delta_{ij}$ the flat spatial metric, but other choices are also
possible, and in this case $K_{ij}^{(0)}=0$, and $\lambda$ an ordering parameter to define
the expansion  
% Modified on 10-10-2005
\cite{wald}.
The ordering parameter $\lambda$ is introduced here just as a computational tool to obtain the
perturbation series. As said in sec.\ref{sec2} this is a standard approach in perturbation theory.
The proper parameter emerges naturally after the series is computed if the series exists and this
is the case as we are going to prove both for $\lambda\rightarrow 0$, the known case, and 
$\lambda\rightarrow\infty$. So,
% end mod 
the equations we get are
\begin{eqnarray}
    \partial_t\gamma_{ij}^{(1)} &=& -2K_{ij}^{(1)} \\ \nonumber 
    \partial_t\gamma_{ij}^{(2)} &=& -2K_{ij}^{(2)} \\ \nonumber
    &\vdots&  \\ \nonumber
    \partial_tK_{ij}^{(1)} &=& -{1\over 2}\Delta_2\gamma_{ij}^{(1)} -{1\over 2}\partial_i\partial_j\gamma^{(1)}
    +{1\over 2}\partial_i\partial^m\gamma_{mj}^{(1)}+{1\over 2}\partial_j\partial^m\gamma_{mi}^{(1)} \\ \nonumber
    \partial_tK_{ij}^{(2)} &=& -{1\over 2}\Delta_2\gamma_{ij}^{(2)} -{1\over 2}\partial_i\partial_j\gamma^{(2)}
    +{1\over 2}\partial_i\partial^m\gamma_{mj}^{(2)}+{1\over 2}\partial_j\partial^m\gamma_{mi}^{(2)} \\ \nonumber
    & &-2K_{il}^{(1)}K_j^{l(1)}+K^{(1)}K_{ij}^{(1)} \\ \nonumber
    & &-{1\over 2}\gamma^{lm(1)}(\partial_l\partial_m\gamma_{ij}^{(1)}+\partial_i\partial_j\gamma_{lm}^{(1)}
    -\partial_i\partial_l\gamma_{mj}^{(1)}-\partial_j\partial_l\gamma_{mi}^{(1)}) \\ \nonumber
    & &-{1\over 2}(\partial_i\gamma_{j}^{p(1)}
    +\partial_j\gamma_{i}^{n(1)}-\partial^p\gamma_{ij}^{(1)})\partial^m\gamma_{mp}^{(1)} \\ \nonumber
    & &-{1\over 2}\partial^m\gamma_{i}^{p(1)}\partial_p\gamma_{jm}^{(1)}
    +{1\over 2}\partial^m\gamma_{i}^{p(1)}\partial_m\gamma_{jp}^{(1)} \\ \nonumber
    & &+{1\over 4}\partial_i\gamma_j^{p(1)}\partial_p\gamma^{(1)}
    +{1\over 4}\partial_j\gamma_i^{p(1)}\partial_p\gamma^{(1)}
    -{1\over 4}\partial^p\gamma_{ij}^{(1)}\partial_p\gamma^{(1)}
    +{1\over 4}\partial_i\gamma_{n}^{m(1)}\partial_j\gamma_m^{n(1)} \\ \nonumber
    &\vdots&
\end{eqnarray}
that as expected display at the leading order the dynamics of gravitational waves. Different choices of the gauge
freedom do not change the physics. The leading order is ruled just by the linear part of Einstein equations.
In this way we recognize the standard weak field expansion \cite{wald} that holds for $\lambda\rightarrow 0$.

We now compute the dual series to the weak field expansion 
and we will show that the situation is quite similar to the two dimensional case
with the notable difference that solutions of Einstein equations have singularities. 
As said, in order to get a non trivial expansion we have to consider the following settings
\begin{eqnarray}
    \tau &=& \sqrt{\lambda}t \\ \nonumber
	  K_{ij} &=& \sqrt{\lambda}\left(K_{ij}^{(0)}+\frac{1}{\lambda}K_{ij}^{(1)}+\frac{1}{\lambda^2}K_{ij}^{(2)}
	  +\ldots\right) \\ \nonumber
	  \gamma_{ij} &=& \gamma_{ij}^{(0)}+\frac{1}{\lambda}\gamma_{ij}^{(1)}+\frac{1}{\lambda^2}\gamma_{ij}^{(2)}
	  +\ldots \\ \nonumber
	  \alpha &=& \alpha_0 + \frac{1}{\lambda}\alpha_1+\frac{1}{\lambda^2}\alpha_2+\ldots
\end{eqnarray}    
while the shift vector $\beta^i$ is taken to be zero. 
The following set of equations is so obtained
\begin{eqnarray}
\label{eq:Set}
    \partial_{\tau}\gamma_{ij}^{(0)} &=& -2\alpha_0K_{ij}^{(0)} \\ \nonumber
    \partial_{\tau}\gamma_{ij}^{(1)} &=& -2\alpha_1K_{ij}^{(0)}-2\alpha_0K_{ij}^{(1)} \\ \nonumber
    &\vdots& \\ \nonumber
    \partial_{\tau} K_{ij}^{(0)} &=& -2\alpha_0K_{il}^{(0)}K_{j}^{l(0)} + \alpha_0K^{(0)}K_{ij}^{(0)} \\ \nonumber
    \partial_{\tau} K_{ij}^{(1)} &=& -2\alpha_1K_{il}^{(0)}K_{j}^{l(0)}-2\alpha_0K_{il}^{(1)}K_{j}^{l(0)}
                                     -2\alpha_0K_{il}^{(0)}K_{j}^{l(1)} \\ \nonumber
                                 & & + \alpha_1K^{(0)}K_{ij}^{(0)} + \alpha_0K^{(1)}K_{ij}^{(0)}   
                                     + \alpha_0K^{(0)}K_{ij}^{(1)} \\ \nonumber
    & &-{1 \over 2}\alpha_0\gamma^{lm(0)}\left\{\partial_l\partial_m\gamma_{ij}^{(0)}
    +\partial_i\partial_j\gamma_{lm}^{(0)}
    -\partial_i\partial_l\gamma_{mj}^{(0)}-\partial_j\partial_l\gamma_{mi}^{(0)}\right. \\ \nonumber
    & & +\gamma^{np(0)}\left[(\partial_i\gamma_{jn}^{(0)}+\partial_j\gamma_{in}^{(0)}
    -\partial_n\gamma_{ij}^{(0)})\partial_l\gamma_{mp}^{(0)}\right. \\ \nonumber
    & & +\partial_l\gamma_{in}^{(0)}\partial_p\gamma_{jm}^{(0)}
    -\partial_l\gamma_{in}^{(0)}\partial_m\gamma_{jp}^{(0)}\left.\right] \\ \nonumber
    & &-{1 \over 2}\gamma^{np(0)}\left[(\partial_i\gamma_{jn}^{(0)}+\partial_j\gamma_{in}^{(0)}
    -\partial_n\gamma_{ij}^{(0)})\partial_p\gamma_{lm}^{(0)}+
    \partial_i\gamma_{ln}^{(0)}\partial_j\gamma_{mp}^{(0)}\left.\right]\right\} \\ \nonumber
    & &-\partial_i\partial_j\alpha_0+{1\over 2}\gamma^{lm(0)}(\partial_i\gamma_{jm}^{(0)}+\partial_j\gamma_{im}^{(0)}
    -\partial_m\gamma_{ij}^{(0)})\partial_l\alpha_0 \\ \nonumber
    &\vdots&
\end{eqnarray}
So, we succeeded in the derivation of a non trivial set of equations for a strong coupling expansion of the Einstein
equation. We see that in this case we have to take $\lambda\rightarrow\infty$ to make the expansion meaningful. This is the main result of the paper that permits us to draw the following conclusions. At the leading order the set of equations that rules the dynamics is that part of the Einstein equations that holds for an homogeneous spacetime. So, unless boundary or initial conditions happen to depend on spatial coordinates, only the time variable appear to be significant in a strong gravitational field. It is interesting to notice here that in the above equations the nonlinear part of Einstein equations has taken the role of the linear part that we have seen to rule the weak field perturbation theory and
a direct evidence of the duality principle in perturbation theory as discussed in sec.\ref{sec2}. The
situation is exactly mirroring the case of the two dimensional gravity by the Liouville equation
with the difference that here we have a singular behavior. Indeed, we
see that due to the peculiar structure of the Einstein equations this should not come out as a surprise. The most
important conclusion to be drawn from this result is that in a strong gravitational field, that is
near a singularity, the spacetime is indeed homogeneous reaching the
conclusion we aimed to. This is the well-known Belinski-Khalatnikov-Lifshitz (BKL)
conjecture \cite{lk,blk1,blk2}.
Indeed , one can complete the proof by assuming $\alpha=\alpha_0=1$ in the identical way
as we did for the weak field perturbation theory. This just simplifies the equations without changing the conclusions
consistently with the duality principle in perturbation theory. Finally we note that the computation of higher
order corrections can be done near a singularity displaying the corrections due to irregularities in the spacetime
as one departs from the singularity when proper boundary conditions are given. 

Two main points should be addressed to complete our proof. Indeed, one should understand how
general are the above equations with respect to a simple rephrasing of the BKL solutions.
But eqs.(\ref{eq:Set}) are able to express more general physical situations as also we will
see in the next section. A typical situation is given by a strongly perturbed black hole
that can be studied. Interesting applications should be also given beyond the close
approximation in collision between black holes \cite{pp,ap}. The other point is the meaning
of the rescaling in time. We know that a small perturbation theory is just a series in $G$,
being $G$ the Newton constant. A dual expansion has meaning as
an expansion of $1/G^n$, being $n$ a positive integer 
as we will see below. In turn, this means that the scaling of
time as $t\sqrt{G^n}$ in this limit implies a prevalence of time variable with respect to
other ones into the evolution equations. This should be compared with the formal
parameter $\lambda$ introduced above.

It is interesting to note that what we have here is a gradient expansion \cite{salo} that now appears
the dual counterpart to the weak expansion for a gravitational field giving us an interesting
interpretation of this expansion as the strong coupling series for general relativity.

\section{Applications\label{sec5a}}

In order to show in a given example that a gradient expansion is indeed a $1/G$ perturbation
series we consider a perturbation on a Schwarzschild black hole and analyze it very
near the horizon of the singularity.

First of all, when one puts ($r_g=2GM$ is the Schwarzschild radius)
\begin{eqnarray}
    \alpha &=& \alpha_0 = 1-\frac{r_g}{r} \\ \nonumber
	\gamma_{rr}^{(0)} &=& -\frac{1}{1-\frac{r_g}{r}} \\ \nonumber
	\gamma_{\theta\theta}^{(0)} = -r^2 & & \gamma_{\phi\phi}^{(0)} = -r^2\sin^2\theta
\end{eqnarray}
into eqs.(\ref{eq:Set}), it is easily verified that higher order corrections are all zero
as it should be as the Schwarzschild solution is exact. This is a very easy consistency check.

Now, we change the physical situation by imposing on the lapse function the choice
\begin{equation}
    \alpha(r,t) = 1-\frac{r_g}{r} + A_0 \frac{r}{r_g}\sin(\omega t)
\end{equation}
that is we perturb the black hole with a radial perturbation of pulsation $\omega$ and
amplitude $A_0$. This is a standard approach to perturbation in general relativity. What
we want to check is the behavior with respect to $G$ of the first order correction.

The solution of eqs.(\ref{eq:Set}) is straightforward to obtain giving the following
solutions
\begin{eqnarray}
    \gamma_{rr} &=& -\frac{1}{1-\frac{r_g}{r}}
	\left[1-A_0^2\frac{1}{rr_g}\frac{3-4\cos{\omega t}+\cos{2\omega t}}{4\omega^2}
	-A_0\left(1-\frac{r_g}{r}\right)\frac{\omega t-\sin{\omega t}}{\omega^2 r^2}+\ldots\right] \\ \nonumber
	\gamma_{\theta\theta} &=& -r^2
	\left[1-2A_0^2\left(1-\frac{r_g}{r}\right)\frac{r}{r_g}\frac{3-4\cos{\omega t}+\cos{2\omega t}}{4\omega^2 r^2}
	-2A_0\left(1-\frac{r_g}{r}\right)^2\frac{r}{r_g}
	\frac{\omega t-\sin{\omega t}}{\omega^2 r^2}+\ldots\right] \\ \nonumber
	\gamma_{\phi\phi} &=& \gamma_{\theta\theta}\sin^2\theta.
\end{eqnarray}
Then, when we approach the horizon of the singularity, that is for $r\approx r_g$, it is easy
to get
\begin{eqnarray}
    \gamma_{rr} &\approx& -\frac{1}{1-\frac{r_g}{r}}
	\left[1-A_0^2\frac{3-4\cos{\omega t}+\cos{2\omega t}}{4\omega^2 r_g^2}+\ldots\right] \\ \nonumber
	\gamma_{\theta\theta} &\approx& -r^2 \\ \nonumber
	\gamma_{\phi\phi} &=& \gamma_{\theta\theta}\sin^2\theta.
\end{eqnarray}
where we recognize that the first order correction goes like $1/\omega^2 r_g^2$ that is, it
is a $1/G^2$ term. This is consistent with our expectations given in the preceding section.

As a final consideration, we know that the gradient expansion at the leading order admits
the cosmological solutions due to Belinski-Khalatnikov-Lifshitz (BKL) \cite{lk,blk1,blk2}.
In view of our analysis, these solutions are a generic effect at the leading order 
of a $1/G$ expansion, and so they appear as a rather general
behavior in such physical situations supporting recent numerical findings \cite{garf}
describing the behavior of solutions of the Einstein equations near a singularity
that seems to be homogeneous in space.

\section{Conclusions\label{sec5}}

We have built an expansion for Einstein equations in vacuum for the description of a strong gravitational field.
The main conclusion is that the spacetime is homogeneous near a singularity.
An open question is left: What is the proper Bianchi solution in the
neighborhood of a singularity? 
The kind of Bianchi spacetime to be considered relies in the end on the proper setting of the initial
conditions. Two other important conclusions to be drawn is that spacetime near the singularity of the big bang
was homogeneous. A future theory of quantum gravity will have the duty to
explain such a behavior near a singular point of the classical theory. 

Further studies are deserved when matter is present. 
In this case the nature of the singularity may change \cite{rend}.

So, we have reached relevant conclusions by the analysis of the Einstein equations in vacuum in a strong coupling 
regime. A lot of physics could be extracted by further analysis of this perturbation equations in 
different situations that at present can be managed only numerically.

\section*{Acknowledgments}

This paper has been inspired by the beautiful lectures on general relativity 
of Vladimir Belinski that I listened at
University of Rome on 1992 where I learned for the first time about BKL
solution and gravitational solitons.

\end{document}